*Original Article*

# Analyzing the Performance of a 2.72kW$_p$ Rooftop Grid-tied Photovoltaic System in Tarlac City, Philippines

Aldrin Joar R. Taduran[1], Leo P. Piao[2]

[1,2]Department of Electrical and Electronics, College of Engineering, Tarlac City State University, Tarlac City, Philippines.

[1]Corresponding Author : ajrtaduran@tsu.edu.ph

***Abstract*** - *Residential and industrial areas are using rooftop grid-tied Photovoltaic (PV) systems, which are becoming increasingly popular. This is because solar energy reduces electrical consumption and provides free energy, while also lowering carbon emissions to create a more sustainable environment. This paper aims to analyze the 2.72kW$_p$ rooftop grid-tied PV system performance between 2020 and 2023 in Tarlac City, Philippines. The PV generated yearly is measured by Array Yield ($Y_A$), Reference Yield ($Y_R$), and Final Yield ($Y_F$), which were found to be valued at 3.12, 3.9, and 3.01 kWh/kW$_P$, respectively. The efficiency can decrease due to System Loss ($L_S$) and Capture Loss ($L_C$), which were 0.78 and 0.12 kWh/kW$_p$, respectively. This results in a Capacity Utilization Factor (CUF) of 15.52% and a Performance Ratio (PR) of 77.10%. The productivity of PV resulted in an array efficiency ($\eta_{array}$) was 12.89%, an inverter efficiency ($\eta_{inv}$) was 94.3%, and a system efficiency ($\eta_{sys}$) was 12.16%. PV energy generation was 3,699 kWh, with 2380 kWh fed into the grid annually. The system's annual revenue is $690.59. The payback period is 6 years with a 238.2% Return On Investment (ROI). Carbon emissions are reduced by 0.379 tCO$_2$/kW$_p$/yr.*

## 1. Introduction

The global electricity demand has increased, and conventional energy sources that rely on fossil fuels are running out. This, combined with environmental issues, leads to the use of Renewable Energy (RE) sources, particularly solar energy [1]. RE is necessary because oil and gas are decreasing, which is a growing problem of global warming [2]. In Southeast Asia, energy demand has rapidly increased by 83% to 701 Mteo from 2000 to 2018 [3]. This is needed in urbanization and economic growth due to the increase in population [4]. The ASEAN (Association of Southeast Asian Nations) established a goal to raise the region's share of RE by 23% by 2025 [5]. The demand for RE will accelerate energy development and carbon emissions in the ASEAN countries. The development will help provide affordable, reliable, clean energy that benefits the communities. The electrical and power industries in the Philippines are experiencing a supply crisis due to high demand. RE installation is a good solution to ensure sufficient sources [6]. As a member of ASEAN, the Philippine government intends to expedite the RE resource transition by permitting foreign ownership of RE projects [7]. The Philippines National Renewable Energy Program (NREP) states that by 2030 and 2040, the country wants to increase its generation by 35% and 50%, respectively [8]. The government wants to increase the general public's access to and affordability of RE. "Renewable Energy Act of 2008" (Republic Act 9513). This law will expedite the development of RE resources in the nation. [9]. The regulation promotes using renewable energy sources as an alternative to electricity, like solar, wind, hydro, biomass, geothermal, and ocean energy [10]. Solar PV systems have the greatest research, development, and cutting-edge technology worldwide among RE sources [11]. The policy encourages the development of solar energy using both off-grid and on-grid solar PV systems. The nation aims to electrify all its urban areas by 2040 to increase the usage of RE through the primary grid, according to the Electrical Power Industry Reform Act (EPIRA), also known as Republic Act No. 9136. [12]. However, certain solar projects usually fail because of limited capacity and financial constraints owing to the delayed implementation of energy laws and policies. The Feed-In Tariff (FIT) and net metering schemes are among the policies.

The net metering schemes for solar PV systems with a generation of less than 100 kW serve as a framework for this paper. The PV excess energy is subtracted from the consumer's monthly bill [13]. This would return the investment with a lower electrical cost. However, RE deployment is often avoided due to policy barriers, environmental concerns, and a lack of knowledge of the economic impact of solar energy [14]. The collaboration between the government and the community will accelerate the nation's solar PV adaptation. More development in the infrastructure and the integration of the solar PV systems into the main grid may result from this. This is essential for the nation's sustainable energy solution [15]. Grid-tied solar PV systems installed on rooftops are common in homes as an alternate power source [16]. Installing a roof-mounted system has significantly contributed to greenhouse gas-free energy generation [17]. The small-scale rooftop grid-tied solar system has huge potential in urban areas with no land cost and decreased transmission and distribution costs. Therefore, there is a significant increase in the number of installations, quality, and performance due to its demand [18]. There is a lot of potential in small rooftop PV systems, which have become more concentrated on research on quality and performance [19].

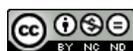

Advanced PV technologies can significantly improve rooftop solar performance. Innovations like bifacial panels, smart inverters, and tracking systems can improve the performance. [20]. Bifacial panels can gain 18-22% efficiency compared to monocrystalline and polycrystalline with 15% to 22% and 15-18%, respectively, because their reflective surface causes a higher yield [21]. Monocrystalline cells are the second most efficient but perform better on small rooftops. However, this research used polycrystalline cells to be cost-effective and suitable for large rooftops [22]. Smart inverters can optimize efficiency by up to 5-20%, which is also highly suitable for any rooftop. Tracking System's typical efficiency gain is 10-25% [23]. While these technologies offer benefits, geographical location must be considered. Proper design is essential to achieve optimal performance.

Many research papers have been presented and published on rooftop grid-tied solar PV systems in different countries. According to the studies, rooftop grid-tied PV systems were conducted at Kumira, Chittagong, for a remote building at a university. The 135-kW solar PV grid-tied system design and modelling has an ideal average solar irradiance of 585.8 W/m$^2$ [24]. Another paper has proposed an alternative approach, which is the implementation of a suitable rooftop PV system in Uganda.

The potential energy yield that may support up to 25.4 MW$_p$ falls between 1046 kWh/kW and 1344 kWh/kW [25]. A similar study about the performance analysis of an 11.2 kW$_p$ rooftop grid-connected PV system in Eastern India. They generated an energy level of 14.96 MWh and a performance ratio of 0.78 [26]. One study in the Philippines, located in San Fernando, Pampanga, evaluated the performance of a 676.8 kW Grid-Tied solar system for commercial buildings. According to the research, the location has shown a solar radiation of 5.50 kWh/m$^2$ with a performance ratio of 0.825 [27].

Numerous studies have been conducted on the performance of rooftop PV in various regions. There is a research gap in medium-sized cities in the Philippines, such as Tarlac City. Existing research has primarily focused on large solar farms, national RE policies, and evaluation in metropolitan areas [28]. Localized, long-term performance data of residential rooftop PV installation where environmental conditions, consumption, and grid infrastructure have different impacts. Addressing this gap in the region will be beneficial for the potential distribution of RE.

This study uses data collected between January 2020 and December 2023 to analyze the 2.72kW$_p$ rooftop grid-tied PV system. The novelty of this study is that it provides details of system performance metrics, economic impact, and environmental impact in Tarlac City. Typically, existing literature often emphasizes short-term assessment or simulation-based methods, while this study used real-time data. This study also looks at the effect of weather conditions throughout the year.

## 2. Methodology
The methodology started with the site location and system description. This was followed by the collection of meteorological data and the calculation of performance metrics. The results were benchmarked against related studies in the various regions, with the expectation that it would give insights into the economic and environmental impact.

### *2.1. Site Location*
A household in Tarlac City had a rooftop PV system with a generating power of 2.72 kWp, an elevation of 10 meters, coordinates of 15.48°N, 120.65°E, and an albedo of 0.2. This geographical advantage enabled the solar panel to absorb more solar irradiance since this system had been dependent on the amount of sunlight received.

### *2.2. System Description*
The grid-tied block diagram is shown in Figure 1. This consisted of a PV module, Micro Circuit Breaker (MCB), Surge Protection Device (SPD), on-grid inverter, and isolator switch. The system description had an installed total rating of 2.72 kW$_p$ and took up 16.1 m$^2$ of space on the rooftop, which was arranged in two 2 x 4 meters arrays. Eight 340W JA solar polysilicon panel modules were installed to provide the total rated power, and a single 3kW grid-tied Solis 5G inverter was slightly oversized.

The plane tilt and the azimuth must be set to 26° and 165°, respectively. The polycrystalline solar panel that was being used had a 340Wp rating. The efficiency of this panel in each cell area was 19.41%. The polysilicon open circuit voltage ($V_{oc}$) and short-circuit current ($I_{sc}$) were 45.9V and 9.43A, respectively. It can operate at a maximum temperature of 25 °C. The rooftop had both free-standing and fixed-mounted solar panels installed. Net metering was an RE incentive mechanism. The rooftop PV system was installed in a typical residential household with one 1.5 conditioning unit.

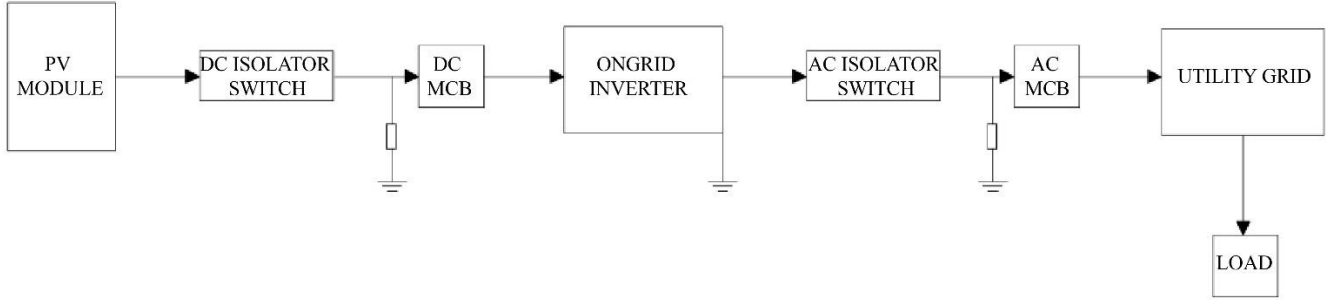

Fig. 1 Block diagram grid-tied

Table 1. Monthly average meteorological data of tarlac city, philippines

|  | Temperature (°C) | Wind speed (m/s) | GHI (W/m$^2$) | DNI (W/m$^2$) | DHI (W/m$^2$) | Plane of Array ($G_{POA}$) irradiance (kWh/m$^2$) |
|---|---|---|---|---|---|---|
| January | 23.49 | 1.77 | 212.85 | 266.46 | 57.86 | 4.73 |
| Febraury | 24.91 | 2.39 | 248.13 | 289.47 | 62.98 | 4.89 |
| March | 26.8 | 1.84 | 286.78 | 324.30 | 60.58 | 5.67 |
| April | 28.99 | 1.90 | 291.33 | 296.88 | 73.12 | 5.81 |
| May | 28.08 | 1.32 | 270.51 | 252.09 | 84.32 | 5.93 |
| June | 27.76 | 1.30 | 275.55 | 253.43 | 88.07 | 5.33 |
| July | 26.78 | 1.24 | 262.17 | 236.57 | 88.80 | 5.05 |
| August | 27.14 | 1.36 | 217.26 | 132.30 | 118.49 | 4.49 |
| September | 26.23 | 1.10 | 250.43 | 219.81 | 89.49 | 4.12 |
| October | 26.14 | 1.47 | 187.39 | 134.76 | 95.54 | 3.85 |
| November | 26.09 | 1.75 | 185.91 | 193.41 | 68.46 | 4.33 |
| December | 25.12 | 1.31 | 170.17 | 181.52 | 65.15 | 3.68 |
| Average | 26.47 | 1.56 | 238.21 | 231.75 | 79.40 | 4.82 |

### 2.3. Data Collection and Monitoring

The Solis 5G inverter sent a data logger through the Soliscloud online platform, which was available in a built-in web browser and mobile application. The solar panel's energy generation and excess per day can be downloaded hourly. The platform can monitor daily temperature, wind speed, weather status, and carbon dioxide reduction. Solar irradiance was obtained from the weather station.

### 2.4. Meteorological Data

The provided meteorological data significantly made an impact on the PV system's performance and efficiency [29]. This was the foundation for estimating solar irradiance in the area or location and evaluating output power. The monthly average temperature, wind speed, Global Horizontal Irradiance (GHI), Direct Normal Irradiance (DNI), Diffuse Horizontal Irradiance (DHI), and Plane Of Array ($G_{POA}$) irradiance were shown in Table 1.

This showed that GHI, DNI, DHI, temperature, and $G_{POA}$ irradiance had the highest value in the months of March to June during the dry season in the Philippines. The months of the wet season, from October to December, had the lowest value.

### 2.5. PV System Performance Metrics

PV system performance metrics are used to compare and assess how well the solar system performs in various climates and locations [30]. The following quantities are used to assess the PV performance:

Array yield ($Y_a$) is measured by the output energy ($E_{DC}$) produced by the array divided by the PV generated ($P_{rated}$), shown in Equation (1) [31].

$$Y_A = \frac{E_{DC}}{P_{rated}} \qquad (1)$$

The energy production under standard test circumstances (STC) divided by its energy production effectively was known as the performance ratio (PR), shown in equation 2 [32]. This was the system's overall effectiveness when compared to its ideal efficiency.

$$PR = \frac{Y_R}{Y_F} \times 100 \qquad (2)$$

The hourly to generate AC output power ($E_{AC}$) on the PV array known as the specific yield ($Y_F$) [33], which was computed by dividing the $P_{rated}$ power of the array shown in Equation (3).

$$Y_F = \frac{E_{AC}}{P_{rated}} \quad (3)$$

Hours needed for a PV array to function under the reference irradiation were known as the reference yield ($Y_R$) [34]. This was measured by the reference irradiance ($G_o$), which is equal to 1000 W/m² divided by the irradiance on a plane ($G_{poa}$) shown in equation 4.

$$Y_R = \frac{G_{poa}}{G_o} \quad (4)$$

Capacity Utilization Factor (CUF) in which the energy generated over that supplied into the grid under maximum rated power is shown in equation 5 [35].

$$CUF = \frac{E_{AC}}{P_{rated} \times 8760} \times 100 \quad (5)$$

The system efficiency ($\eta_{sys}$) is the productivity of the inverter and PV array [36]. This computed $E_{AC}$ is obtained by the $G_{poa}$ times the surface area ($A_g$) of the PV array shown in equation 6.

$$\eta_{sys} = \frac{E_{AC}}{G_{poa} \times A_g} \times 100 \quad (7)$$

The capture loss ($L_c$) was calculated by the difference between $Y_R$ and $Y_A$, shown in equation 7. These losses occur because there is a rise in temperature in the PV cell, dust accumulation, partial shading, and inhomogeneous irradiance [37].

$$L_c = Y_r - Y_a \quad (8)$$

System Loss ($L_s$) was due to wiring resistance, power dissipation in semiconductors, and aging. This was measured by $Y_A$ minus $Y_F$, shown in equation 8.

$$L_s = Y_a - Y_f$$

$$L_s = Y_a - Y_f$$

### 2.6. Economic and Environmental Impact

The sustainability of the 2.72 kW$_p$ rooftop PV system was evaluated through the following: Return On Investment (ROI), Net Present Cost (NPC), Levelized Cost Of Energy (LCOE), and payback period. These were used to evaluate the economic impact. The carbon ($CO_2$) emission balance was used to assess the environmental impact.

## 3. Results and Discussion
### 3.1. Technical Analysis of 2.72 kW$_p$ Rooftop PV System

This discusses the system's performance metric results. The average yield of the $Y_A$, $Y_R$, and $Y_F$ was shown in Table 2. The results were 3.12, 3.9, and 3.01 kWh/kW$_p$, respectively. It was observed that April had the highest PV generation during the dry season. The lowest PV generation was recorded during the wet season in December due to low temperature and solar irradiance. The $L_c$ and $L_S$ were 0.78 and 0.12 kWh/kW$_p$ each year due to dust and dirt, shading, losses in the cable, and mismatch.

The weather conditions highly affected the output PV generated. This was observed by collecting the hourly rate of PV generated and the $G_{poa}$. It was also observed that the difference in generation was due to fluctuation in solar irradiance under the influence of different weather conditions. Correlation analysis was used to evaluate the relationship between the PV generated and solar irradiance under weather conditions.

Table 2. Monthly average energy yield and losses of the system

|  | Cell Temperature (°C) | $E_{AC}$ (kWh) | $E_{DC}$ (kWh) | $Y_A$ (kWh/kW) | $Y_R$ (kWh/m²) | $Y_F$ (kWh/kW$_p$) | $L_C$ (kWh/kW$_p$) | $L_S$ (kWh/kW$_p$) |
|---|---|---|---|---|---|---|---|---|
| January | 34.71 | 5.58 | 5.85 | 2.15 | 2.78 | 2.05 | 0.63 | 0.1 |
| Febraury | 39.83 | 7.94 | 8.24 | 3.03 | 3.86 | 2.92 | 0.83 | 0.11 |
| March | 44.48 | 9.74 | 10.06 | 3.7 | 4.64 | 3.58 | 0.94 | 0.12 |
| April | 48.94 | 12.00 | 12.40 | 4.56 | 5.74 | 4.41 | 1.18 | 0.15 |
| May | 45.09 | 10.47 | 10.85 | 3.99 | 4.9 | 3.85 | 0.91 | 0.13 |
| June | 45.14 | 10.83 | 11.21 | 4.12 | 5.04 | 3.98 | 0.92 | 0.14 |
| July | 39.86 | 8.30 | 8.62 | 3.17 | 3.84 | 3.05 | 0.67 | 0.12 |
| August | 41.8 | 9.19 | 9.55 | 3.51 | 4.29 | 3.38 | 0.78 | 0.13 |
| September | 36.44 | 6.20 | 6.47 | 2.38 | 2.91 | 2.28 | 0.53 | 0.11 |
| October | 37.68 | 6.83 | 7.13 | 2.62 | 3.26 | 2.51 | 0.64 | 0.11 |
| November | 38.96 | 6.28 | 6.56 | 2.41 | 3.14 | 2.31 | 0.72 | 0.1 |
| December | 35.17 | 4.81 | 5.06 | 1.86 | 2.45 | 1.77 | 0.58 | 0.09 |
| **Average** | **40.8** | **8.18** | **8.50** | **3.12** | **3.9** | **3.01** | **0.78** | **0.12** |

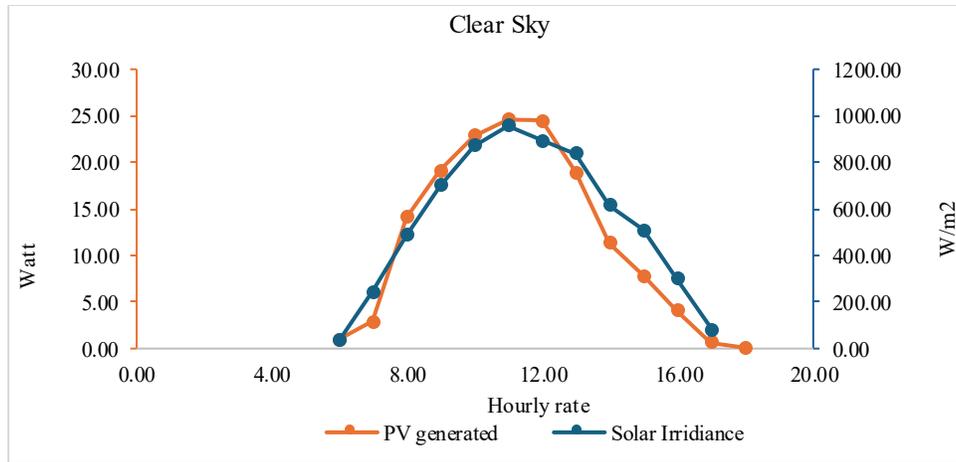

**Fig. 2 Clear sky hourly rate of PV Generated and Solar irradiance**

In Figure 2, the average output on a clear sky was 14.8 kWh. This showed that a clear sky has a 0.784 strong positive correlation, and the system can achieve an optimal PV generation with high solar irradiance. In Figure 3, the partially cloudy conditions had PV outputs of an average of 11.9 kWh and a 0.728 strong positive correlation. It was a reliable output even with the partial shading in the solar panels. In Figure 4, the overcast condition had a lower PV output of 9.2 kWh. It had a value of 0.636, a moderately positive correlation due to the total shading.

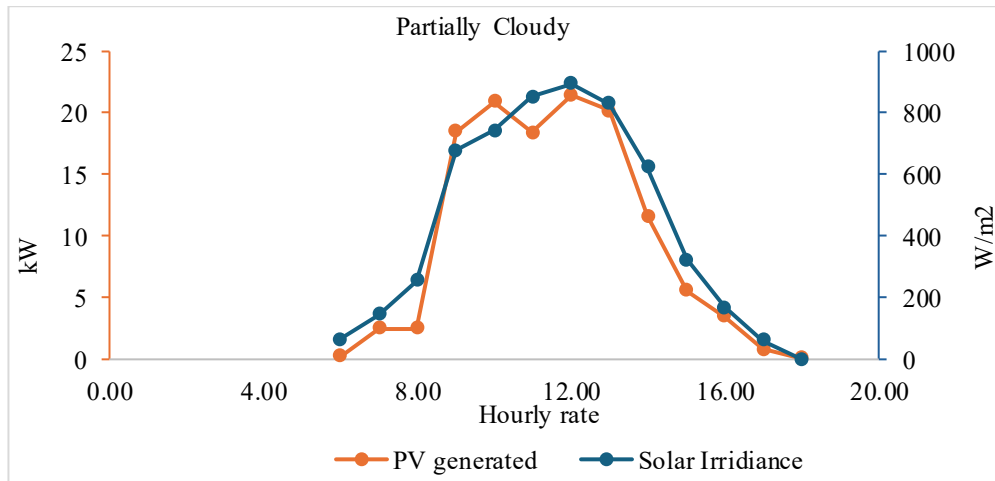

**Fig. 3 Partially Cloudy Hourly Rate of PV Generated and Solar Irradiance**

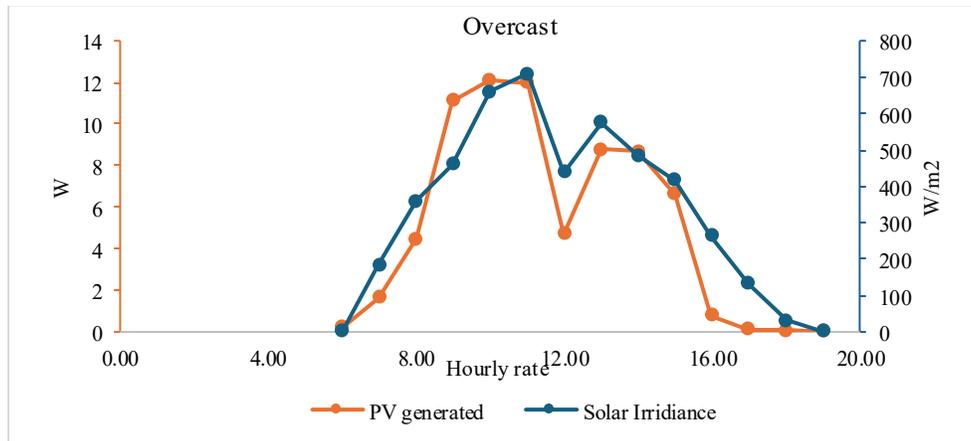

**Fig. 4 Overcast Hourly Rate of PV Generated and Solar Irradiance**

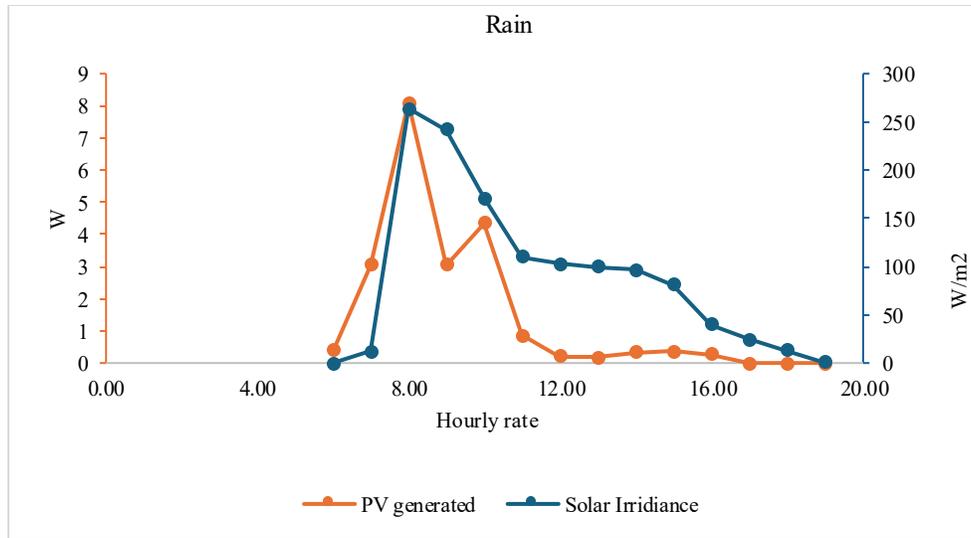

**Fig. 5 Rain Hourly Rate of PV Generated and Solar Irradiance**

In Figure 5, the output was 2.1 kWh during the rainy weather conditions. The correlation had a weak positive correlation of 0.445. This caused the output generation to drop as the solar irradiance decreased.

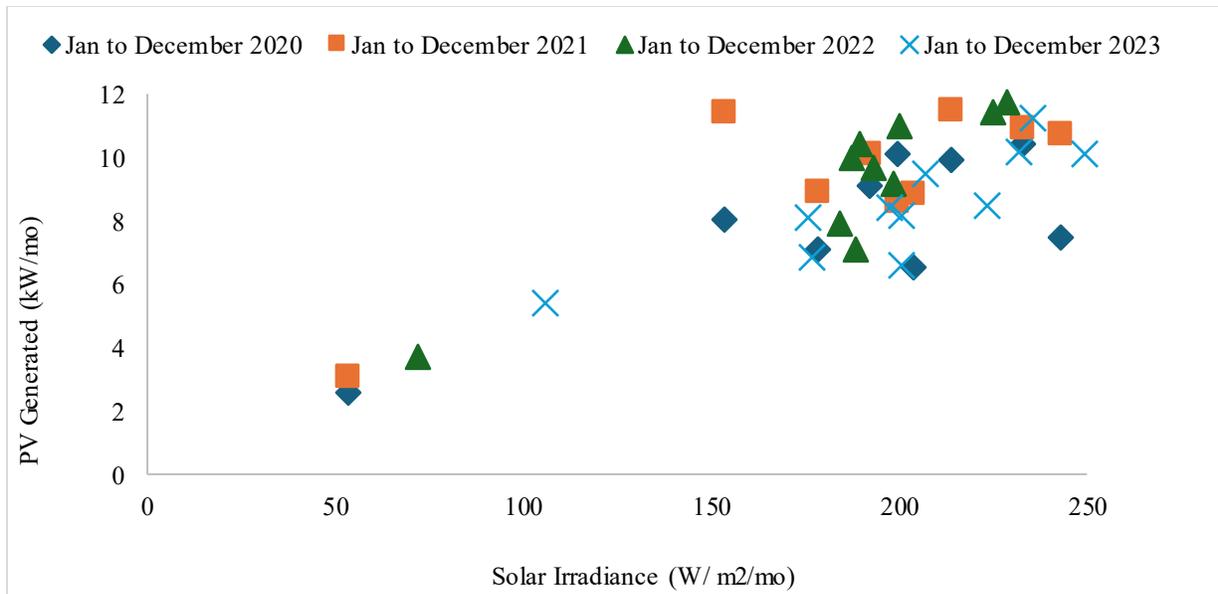

**Fig. 6 Correlation of PV generation versus Solar Irradiance from January 2020 to December 2023**

The relationship between daily PV generation and daily solar irradiation between January 2020 and December 2023 is shown in Figure 6. It had a moderately positive correlation of 0.679 from 2020 to 2023. Other factors influencing the relationship between PV generation and solar irradiance included the wind, cell temperature, dust, shading, array degradation, and system losses. This indicated that the solar irradiance, ambient temperature, and cloud cover remained the factors influencing the system output. This validated that the performance was largely dependent on the solar irradiance for the basis of yield prediction and capacity factor estimation. The monthly average of CUF, PR, $\eta_{array}$, $\eta_{inv}$, $\eta_{sys}$, and energy supplied to the grid were used to analyze the system's efficiency, which was shown in Table 3. This measurement was based on the region, GHI of the PV location, $\eta_{inv}$, and $\eta_{array}$. The PR varied from 60% to 80%, compared to the present study PR with a value of 77.10%. This indicated a good performance for a rooftop PV system. The CUF varied between 15.58% and 15.12% from March to April, with an average of 15.52% annually. The annual averages for $\eta_{inv}$ and $\eta_{array}$ were 95.8% and 13.57%, respectively. The current system efficiency was 13%, indicating that the inverter and array were performing satisfactorily. The energy supplied to the grid generated 2380 kWh per year.

Table 3. Monthly Average Energy supplied to Grid, Array Efficiency, Inverter Efficiency, CUF, PR, and System Efficiency

| | Energy Supplied to Grid (KWh) | η$_{array}$ | η$_{inv}$ | CUF | PR | η$_{sys}$ |
|---|---|---|---|---|---|---|
| January | 159.75 | 13.09% | 94.7% | 11.78% | 73.60% | 12.40% |
| February | 186.75 | 13.30% | 95.8% | 12.36% | 75.50% | 12.74% |
| March | 183.00 | 13.52% | 96.3% | 15.58% | 77.10% | 13.02% |
| April | 202.00 | 13.47% | 96.5% | 15.57% | 77.00% | 13.00% |
| May | 226.75 | 13.79% | 96.3% | 15.12% | 78.70% | 13.28% |
| June | 229.50 | 13.84% | 96.4% | 13.32% | 79.00% | 13.34% |
| July | 210.50 | 14.00% | 95.8% | 12.03% | 79.50% | 13.41% |
| August | 198.00 | 13.86% | 96% | 12.33% | 78.80% | 13.31% |
| September | 206.5 | 13.87% | 95% | 12.57% | 78.20% | 13.18% |
| October | 191.00 | 13.61% | 95.3% | 12.03% | 76.90% | 12.97% |
| November | 194.50 | 13.03% | 95.2% | 10.82% | 73.70% | 12.40% |
| December | 191.750 | 12.89% | 94.3% | 11.75% | 72.40% | 12.16% |

Table 4. PV Performance Summary of Researcher Related to Rooftop Grid-Tied System.

| Country/ Site Location | System Capacity | PR | CUF | Module Type | η$_{sys}$ | References |
|---|---|---|---|---|---|---|
| Kuala Terengganu, Malaysia | 7.8 kW$_p$ | 75.72% | 13-16% | mono-Si | 10%-12% | [38] |
| Jakarta, Indonesia | 5 kW$_p$ | 76.07% | 11.13% | poly-Si | - | [39] |
| Hue, Vietnam | 1.32 kW$_p$ | 78.11% | 15.07% | poly-Si | 12.89% | [40] |
| Cebu, Philippines | 8.36 kW$_p$ | 40.1%-77.8% | 18.96% | poly-Si | - | [41] |
| Tak province, Thailand | 3.5 kWp | 59%-76.4% | - | poly-Si | - | [42] |
| Northern India | 5 kW$_p$ | 76.97% | 16.39% | poly-Si | 10.02% | [43] |
| Male, Maldives | 6.6 kW$_p$ | 81.56% | 18.89% | poly-Si | 13.87% | [44] |
| Central Java, Indonesia | 30 kW$_p$ | 79.40% | - | poly-Si | - | [45] |
| Mae Hong Son, Thailand | 11 kW$_p$ | 73.45% | 14% | poly-Si/mono-Si | 10.41% | [46] |
| Singapore | 142.5 kW$_p$ | 81.00% | 15.70% | poly-Si | 11.20% | [47] |
| Norway | 2.07 kW$_p$ | 83.03% | 10.58% | poly-Si/mono-Si | 11.60% | [48] |
| Port Elizabeth, South Africa | 3.2 kW$_p$ | 64.30% | 20.41% | poly-Si | - | [49] |
| Turkey | 2.73 kW$_p$ | 72% | 15.69% | poly-Si | - | [50] |
| Ireland | 1.72 kW$_p$ | 81.50% | 10.10% | mono-Si | 13.30% | [51] |
| Present Study | 2.72 kW$_p$ | 77.10% | 15.52% | poly-Si | 13.00% | |

Table 4 shows the summary of the PV performance of the research study related to the rooftop grid-tied PV system in different locations compared to the present study. These results of PR, CUF, and system efficiency in various countries for a small-scale rooftop PV system were used as the basis for comparison. It also showed that the present study had a higher PR compared to the related research in Malaysia, Indonesia, India, South Africa, Thailand, and Turkey due to their GHI locations. In addition, other countries like the Maldives, South Africa, and India had a higher CUF that varied between 16% and 20%, while the present study was equal to 15.51%. The η$_{sys}$ in the present study had the same result in Vietnam, the Maldives, and Ireland.

### 3.2. Economic and Environmental Impact

The economic impact was provided by the Net Present Value (NPV), Levelized Cost Of Energy (LCOE), payback period, and Return On Investment (ROI) for 20 years. From 2020 to 2023, the PV generated was 3699 kWh, and the energy supplied to the grid was 2380 kWh. The NPV calculated the overall cost of the system over its lifetime. NPV was $4,197.26, which will be the system's lifetime expenses.

The LCOE was the overall energy generation cost over the lifetime of the system. LCOE had a value of $0.088 per kWh. The total investment was $1762.12, and the operating cost was PHP 176.21/year for the 2.72 kW$_p$ rooftop PV system. The ROI was 238.2% profitable, and the payback was 6 years. Due to the lower middle income in the Philippines, the net-metering was typically compensated by the average generation cost and a lack of incentive for RE. The average monthly energy supply to the grid and the associated cost savings were shown in Figure 7. The data displayed the seasonal pattern over time of the profit. This demonstrated that the average monthly energy supply to the grid was 198.33 kWh, resulting in a monthly savings of $57.55.

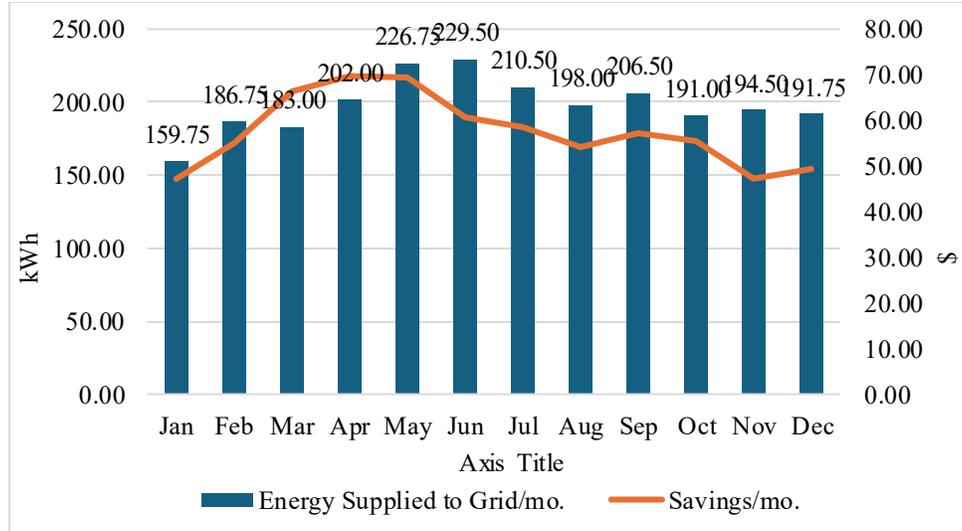

**Fig. 7 Energy supplied to grid and monthly savings**

Energy generation on the PV system was expected to cause a significant impact on the environment for the project's lifetime of 20 years. The Life Cycle Emission (LCE) system was the quantity of carbon emissions produced during the development and manufacturing of the PV system.

The LCE grid was defined by carbon emissions from the grid. The LCE grid and LCE system had a value of 480 $gCO_2/kWh$ and 5.4 $tCO_2$, respectively. It was estimated that the reduced carbon emission into the atmosphere is around 0.379 $tCO_2/kW_p/yr$.

## 4. Conclusion

The 2.72 $kW_p$ rooftop PV system in Tarlac City, Philippines, was successfully presented on the 4 years of PV performance analysis. The effect of weather variation on solar irradiance was converted to PV generation. It was observed that the clear sky and the partially cloudy conditions had a very strong correlation due to the fewer factors affecting the PV array's generation. Meanwhile, the overcast and rainy conditions had a weak correlation due to various factors that can affect the system. This showed a positive correlation between solar irradiance and PV-generated energy for long-term efficiency. The PR, CUF, $\eta_{array}$, $\eta_{inv}$, and $\eta_{sys}$ indicated that the system in Tarlac City, Philippines, produced satisfactory results while operating within the standard parameter set by the other countries.

The economic impact showed a viable financial return with earnings after six years, and a significant impact on reducing $CO_2$ in the environment, which can contribute to renewable energy and climate change. Overall, these provided a practical insight into policies, benefits, and challenges in tropical climates. Future research may expand the scope by applying real-time monitoring, predictive modeling, and a large-scale system for nationwide adaptation.

## Funding Statement
Funding for publication will be granted by the university after it is accepted in a journal.

## Acknowledgments
The authors thank the University Research Office at Tarlac City State University for allowing them to conduct this study. Particularly, special acknowledgement is also extended to the residence's owner.

## Reference

[1] Rohith Goura, "Analyzing the on-Field Performance of a 1-Megawatt-Grid-Tied PV System in South India," *International Journal of Sustainable Energy*, vol. 34, no. 1, pp. 1-9, 2015. [CrossRef] [Google Scholar] [Publisher Link]

[2] Phebe Asantewaa Owusu, and Samuel Asumadu-Sarkodie, "A Review of Renewable Energy Sources, Sustainability Issues and Climate Change Mitigation," *Cogent Engineering*, vol. 3, no. 1, pp. 1-14, 2016. [CrossRef] [Google Scholar] [Publisher Link]

[3] J. Aleluia et al., "Accelerating a Clean Energy Transition in Southeast Asia: Role of Governments and Public Policy," *Renewable and Sustainable Energy Reviews*, vol. 159, 2022. [CrossRef] [Google Scholar] [Publisher Link]

[4] Ravi Ramamurti, "Can Governments Make Credible Promises? Insights from Infrastructure Projects in Emerging Economies," *Journal of International Management*, vol. 9, no. 3, pp. 253-269, 2003. [CrossRef] [Google Scholar] [Publisher Link]

[5] Dolf Gielen et al., "The Role of Renewable Energy in the Global Energy Transformation," *Energy Strategy Reviews*, vol. 24, pp. 38-50, 2019. [CrossRef] [Google Scholar] [Publisher Link]

[6] Feed-in Tariff Scheme in the Philippines: An Overview, Power Philippines, 2017. [Online]. Available: https://powerphilippines.com/feed-in-tariff-scheme-in-the-philippines-an-overview/
[7] Alexander Chipman Koty, Philippines Opens Renewable Energy to Full Foreign Ownership, ASEAN Briefing, 2023. [Online]. Available: https://www.aseanbriefing.com/news/philippines-opens-renewable-energy-to-full-foreign-ownership/
[8] Sahara Piang Brahim, "Renewable Energy and Energy Security in the Philippines," *Energy Procedia*, vol. 52, pp. 480-486, 2014. [CrossRef] [Google Scholar] [Publisher Link]
[9] Republic Act No. 9513, Renewable Energy Act of 2008, Official Gazette of the Republic of the Philippines, 2008. [Online]. Available: https://www.officialgazette.gov.ph/2008/12/16/republic-act-no-9513/
[10] Albert P. Aquino, and Christian L. Albelda, "Renewable Energy Act for Energy Self-Sufficiency and Harmful Emission Reduction," *FFTC Agricultural Policy Platform (FFTC-AP)*, 2014. [Google Scholar] [Publisher Link]
[11] Henrik te Heesen, Volker Herbort, and Martin Rumpler, "Performance of Roof-Top PV Systems in Germany from 2012 to 2018," *Solar Energy*, vol. 194, pp. 128-135, 2019. [CrossRef] [Google Scholar] [Publisher Link]
[12] M. Adoracion Navarro, P. Kristina Ma Ortiz, and Jethro El L. Camara, "How Energy Secure is the Philippines?," *Philippine Institute for Development Studies (PIDS)*, pp. 1-67, 2023. [Google Scholar] [Publisher Link]
[13] Ana Paula Farias-Rocha et al., "Solar Photovoltaic Policy Review and Economic Analysis for on-Grid Residential Installations in the Philippines," *Journal of Cleaner Production*, vol. 223, pp. 45-56, 2019. [CrossRef] [Google Scholar] [Publisher Link]
[14] Emrah Karakaya, and Pranpreya Sriwannawit, "Barriers to the Adoption of Photovoltaic Systems: The State of the Art," *Renewable and Sustainable Energy Reviews*, vol. 49, pp. 60-66, 2015. [CrossRef] [Google Scholar] [Publisher Link]
[15] Joe Zaldarriaga, Powering a Brighter, Sustainable Future with Solar, Philippine News Agency, 2023. [Online]. Available: https://www.pna.gov.ph/opinion/pieces/769-powering-a-brighter-sustainable-future-with-solar.
[16] Chandrakant Dondariya et al., "Performance Simulation of Grid-Connected Rooftop Solar PV System for Small Households: A Case Study of Ujjain, India," *Energy Reports*, vol. 4, pp. 546-553, 2018. [CrossRef] [Google Scholar] [Publisher Link]
[17] Malene Eldegard Leirpoll et al., "Optimal Combination of Bioenergy and Solar Photovoltaic for Renewable Energy Production on Abandoned Cropland," *Renewable Energy*, vol. 168, pp. 45-56, 2021. [CrossRef] [Google Scholar] [Publisher Link]
[18] Wanlin Chen, Shiyu Yang, and Joseph H.K. Lai, "Carbon Offset Potential of Rooftop Photovoltaic Systems in China," *Solar Energy*, vol. 274, 2024. [CrossRef] [Google Scholar] [Publisher Link]
[19] Sven Killinger et al., "On the Search for Representative Characteristics of PV Systems: Data Collection and Analysis of PV System Azimuth, Tilt, Capacity, Yield and Shading," *Solar Energy*, vol. 173, pp. 1087-1106, 2018. [CrossRef] [Google Scholar] [Publisher Link]
[20] Mahesh Raj Nagaraja, Wahidul K. Biswas, and Chithirai Pon Selvan, "Advancements and Challenges in Solar Photovoltaic Technologies: Enhancing Technical Performance for Sustainable Clean Energy - A Review," *Solar Energy Advances*, vol. 5, pp. 1-20, 2025. [CrossRef] [Google Scholar] [Publisher Link]
[21] Marzia Alam, Mehreen Saleem Gul, and Tariq Muneer, "Performance Analysis and Comparison Between Bifacial and Monofacial Solar Photovoltaic at Various Ground Albedo Conditions," *Renewable Energy Focus*, vol. 44, pp. 295-316, 2023. [CrossRef] [Google Scholar] [Publisher Link]
[22] Sugianto, "Comparative Analysis of Solar Cell Efficiency between Monocrystalline and Polycrystalline," *INTEK Journal Penelitian*, vol. 7, no. 2, pp. 92-100, 2020. [CrossRef] [Publisher Link]
[23] Improving Grid Stability with Smart Inverter Technology: The Path to Sustainable Energy, Elege New Energy Manufacturer, 2024. [Online]. Available: https://energy-elege.com/high-efficiency-solar-inverter-solutions-top-rated-technology/
[24] Umme Riazul Jannat Eiva et al., "Design, Performance, and Techno-Economic Analysis of a Rooftop Grid-Tied PV System for a Remotely Located Building," *IET Renewable Power Generation*, vol. 19, no. 1, pp. 1-17, 2023. [CrossRef] [Google Scholar] [Publisher Link]
[25] Nicholas Mukisa, Ramon Zamora, and Tek Tjing Lie, "Feasibility Assessment of Grid-Tied Rooftop Solar Photovoltaic Systems for Industrial Sector Application in Uganda," *Sustainable Energy Technologies and Assessments*, vol. 32, pp. 83-91, 2019. [CrossRef] [Google Scholar] [Publisher Link]
[26] Renu Sharma, and Sonali Goel, "Performance Analysis of a 11.2 kWp Roof Top Grid-Connected PV System in Eastern India," *Energy Reports*, vol. 3, pp. 76-84, 2017. [CrossRef] [Google Scholar] [Publisher Link]
[27] T Eris Jeremiah Addun et al., "Performance Assessment of 676.8 kW Grid-Tied Solar Power Generating System at S&R San Fernando, Pampanga," *Iconic Research and Engineering Journals*, vol. 6, no. 1, pp. 390-399, 2022. [Publisher Link]
[28] A. D. de Luna et al., "Cost-Benefit Analysis of Converting Agricultural Land into Solar Farm Using RS & GIS: Case of Tarlac Province," *The International Archives of the Photogrammetry, Remote Sensing and Spatial Information Sciences*, vol. XLVI-4/W6-2021, pp. 133-140, 2021. [CrossRef] [Google Scholar] [Publisher Link]
[29] Tarek AlSkaif et al., "A Systematic Analysis of Meteorological Variables for PV Output Power Estimation," *Renewable Energy*, vol. 153, pp. 12-22, 2020. [CrossRef] [Google Scholar] [Publisher Link]

[30] David King, William E. Boyson, and Jay Kratochvil, "Analysis of Factors Influencing the Annual Energy Production of Photovoltaic Systems," *Conference Record of the Twenty-Ninth IEEE Photovoltaic Specialists Conference,* 2002, New Orleans, LA, USA, pp. 1356-1361, 2003. [CrossRef] [Google Scholar] [Publisher Link]

[31] Lena D. Mensah, John O. Yamoah, and Muyiwa S. Adaramola, "Performance Evaluation of a Utility-Scale Grid-Tied Solar Photovoltaic (PV) Installation in Ghana," *Energy for Sustainable Development*, vol. 48, pp. 82-87, 2018. [CrossRef] [Google Scholar] [Publisher Link]

[32] Ishan Purohit, Pallav Purohit, and Shashaank Shekhar, "Evaluating the Potential of Concentrating Solar Power Generation in Northwestern India," *Energy Policy*, vol. 62, pp. 157-175, 2013. [CrossRef] [Google Scholar] [Publisher Link]

[33] M. Malvoni et al., "Long Term Performance, Losses and Efficiency Analysis of a 960kW$_P$ Photovoltaic System in the Mediterranean Climate," *Energy Conversion and Management*, vol. 145, pp. 169-181, 2017. [CrossRef] [Google Scholar] [Publisher Link]

[34] Nallapaneni Manoj Kumar et al., "Performance Analysis of 100 kWp Grid Connected Si-Poly Photovoltaic System using PVsyst Simulation Tool," *Energy Procedia*, vol. 117, pp. 180-189, 2017. [CrossRef] [Google Scholar] [Publisher Link]

[35] K. P. Vasudev et al., "Performance Analysis of a 48 kWp Grid connected Rooftop Photovoltaic System," *2018 4$^{th}$ International Conference for Convergence in Technology (I2CT)*, Mangalore, India, pp. 1-6, 2018. [CrossRef] [Google Scholar] [Publisher Link]

[36] Dragana D. Milosavljević, Tomislav M. Pavlović, and Danica S. Piršl, "Performance Analysis of a Grid-Connected Solar PV Plant in Niš, Republic of Serbia," *Renewable and Sustainable Energy Reviews*, vol. 44, pp. 423-435, 2015. [CrossRef] [Google Scholar] [Publisher Link]

[37] Divine Atsu, Istvan Seres, and Istvan Farkas, "The State of Solar PV and Performance Analysis of Different PV Technologies Grid-Connected Installations in Hungary," *Renewable and Sustainable Energy Reviews*, vol. 141, pp. 1-9, 2021. [CrossRef] [Google Scholar] [Publisher Link]

[38] N. Anang et al., "Performance Analysis of a Grid-Connected Rooftop Solar PV System in Kuala Terengganu, Malaysia," *Energy and Buildings*, vol. 248, 2021. [CrossRef] [Google Scholar] [Publisher Link]

[39] Budiman Kamil et al., "Performance Analysis of Multi-Oriented Residential Rooftop PV System in Indonesia towards Net Zero Emission by 2060," *Progress in Solar Energy and Engineering Systems*, vol. 6, no. 1, pp. 16-22, 2022. [CrossRef] [Google Scholar] [Publisher Link]

[40] Xuan Cuong Ngo, Thi Hong Nguyen, and Nhu Y Do, "A Comprehensive Assessment of a Rooftop Grid-Connected Photovoltaic System: A Case Study for Central Vietnam," *International Energy Journal*, vol. 22, no. 1, pp. 13-24, 2022. [Google Scholar] [Publisher Link]

[41] Khrisydel Rhea M. Supapo, Lorafe Lozano, and Edward M. Querikiol, "Performance Evaluation of an Existing Renewable Energy System at Gilutongan Island, Cebu, Philippines," *Journal of Engineering*, vol. 2024, pp. 1-19, 2024. [CrossRef] [Google Scholar] [Publisher Link]

[42] A. Chaita and J. Kluabwang, "Performance Evaluation of 3.5 kWp Rooftop Solar PV Plant in Thailand," *Proceedings of the International MultiConference of Engineers and Computer Scientists*, vol. 2, 2016. [Google Scholar] [Publisher Link]

[43] Satish Kumar Yadav, and Usha Bajpai, "Performance Evaluation of a Rooftop Solar Photovoltaic Power Plant in Northern India," *Energy for Sustainable Development*, vol. 43, pp. 130-138, 2018. [CrossRef] [Google Scholar] [Publisher Link]

[44] Khalid Mohamed et al., "Operational Performance Assessment of Rooftop PV Systems in the Maldives," *Energy Reports*, vol. 11, pp. 2592-2607, 2024. [CrossRef] [Google Scholar] [Publisher Link]

[45] Andrian Mayka Ariawan, Jaka Windarta, and Sujarwanto Dwiatmoko, "Rooftop PV Plant Development Planning at the Central Java Provincial DPRD Secretariat Office," *Journal of Industrial Pollution Prevention Technology Research*, vol. 13, no. 1, pp. 43-52, 2022. [CrossRef] [Google Scholar] [Publisher Link]

[46] Somchai Chokmaviroj, Rakwichian Wattanapong, and Yammen Suchart, "Performance of a 500kW$_P$ Grid Connected Photovoltaic System at Mae Hong Son Province, Thailand," *Renewable Energy*, vol. 31, no. 1, pp. 19-28, 2006. [CrossRef] [Google Scholar] [Publisher Link]

[47] Stephen Wittkopf et al., "Analytical Performance Monitoring of a 142.5kw$_p$ Grid-Connected Rooftop BIPV System in Singapore," *Renewable Energy*, vol. 47, pp. 9-20, 2012. [CrossRef] [Google Scholar] [Publisher Link]

[48] Muyiwa S. Adaramola, and Emil E.T. Vågnes, "Preliminary Assessment of a Small-Scale Rooftop PV-Grid Tied in Norwegian Climatic Conditions," *Energy Conversion and Management*, vol. 90, pp. 458-465, 2014. [CrossRef] [Google Scholar] [Publisher Link]

[49] D. Okello, E.E. Van Dyk, and F.J. Vorster, "Analysis of Measured and Simulated Performance Data of a 3.2kwp Grid-Connected PV System in Port Elizabeth, South Africa," *Energy Conversion and Management*, vol. 100, pp. 10-15, 2015. [CrossRef] [Google Scholar] [Publisher Link]

[50] Rustu Eke, and Huseyin Demircan, "Performance Analysis of a Multi Crystalline Si Photovoltaic Module under Mugla Climatic Conditions in Turkey," *Energy Conversion and Management*, vol. 65, pp. 580-586, 2012. [CrossRef] [Google Scholar] [Publisher Link]

[51] L.M. Ayompe et al., "Measured Performance of a 1.72kw Rooftop Grid Connected Photovoltaic System in Ireland," *Energy Conversion and Management*, vol. 52, no. 2, pp. 816-825, 2011. [CrossRef] [Google Scholar] [Publisher Link]